\documentclass[twocolumn,showpacs,preprintnumbers,amsmath,amssymb,prb,superscriptaddress]{revtex4}
\usepackage{mathrsfs}
\usepackage{graphicx}
\usepackage{dcolumn}
\usepackage{bm}
\usepackage{amsmath}
\usepackage{amsfonts}
\usepackage{color}

\begin{document}
\title{Weyl Semimetal and Topological Phase Transition in Five Dimensions}
\author{Biao Lian}
\affiliation{Department of Physics, McCullough Building, Stanford University, Stanford, California 94305-4045, USA}
\author{Shou-Cheng Zhang}
\affiliation{Department of Physics, McCullough Building, Stanford University, Stanford, California 94305-4045, USA}

\begin{abstract}
We study two Weyl semimetal generalizations in five dimensions (5d) which have Yang monopoles and linked Weyl surfaces in the Brillouin zone, respectively, and carry the second Chern number as a topological number. In particular, we show a Yang monopole naturally reduces to a Hopf link of two Weyl surfaces when the $\mathbf{TP}$ (time-reversal combined with space-inversion) symmetry is broken. We then examine the phase transition between insulators with different topological numbers in 5d. In analogy to the 3d case, 5d Weyl semimetals emerge as intermediate phases during the topological phase transition.
\end{abstract}

\date{\today}


\maketitle

\section{Introduction}

The discovery of topological states of matter has greatly enriched the variety of condensed matter in nature \cite{Qi2011}. These states usually undergo phase transitions involving a change of topology of the ground state wave function, which are called topological phase transitions (TPTs).
In three dimensions (3d), an significant topological state is the Weyl semimetal \cite{Berry1984,Nielsen1983,Wan2011,Balents2011}, which plays a key role in TPTs of 3d insulators.
An example is the time-reversal invariant (TRI) transition between a noncentrosymmetric topological insulator (TI) \cite{Fu2007,Qi2008} and a normal insulator (NI) in 3d, during which an intermediate TRI Weyl semimetal phase inevitably occurs \cite{Murakami2007,Murakami2016}. Another example is the TPT between different 3d Chern insulators (CI) \cite{Kohmoto1992}, where an intermediate Weyl semimetal phase is also required \cite{Haldane2004}. In both examples, the topological numbers of the insulators are transferred via Weyl points of the Weyl semimetal phase, which behave as "Dirac monopoles" of the Berry curvature in the Brillouin zone (BZ). The electrons around each Weyl point obey the Weyl equation, with a chirality equal to the first Chern number $C_1=\pm1$ of the Berry curvature around the Weyl point.

Recently, there has been a revival of interest in gapless topological phases in higher dimensions, aimed at understanding roles of higher-dimensional topological numbers \cite{Qi2008,Horava2005,Zhao2013,Schnyder2015,Lian2016,Mathai2017,Mathai2016b,Sugawa2016}. In particular, the Weyl semimetal can be generalized to 5d in two ways: the first is to promote Weyl fermions in 3d to chiral fermions in 5d, which are described by a 4-component spinor and have a 2-fold degenerate linear energy spectrum. The Dirac monopoles associated with the Weyl points in 3d become the Yang monopoles in 5d \cite{Yang1978}, which carry a non-Abelian second Chern number $C_2^{NA}=\pm1$ of the SU($2$) Berry curvature of the 2-fold degenerate valence (conduction) band \cite{Wilczek1984}. The Yang monopole was first introduced into condensed matter physics in the construction of the four dimensional quantum Hall effect \cite{Zhang2001}. The second way is to keep the energy spectrum non-degenerate, while promoting the Weyl points to linked 2d Weyl surfaces in the 5d BZ \cite{Horava2005,Lian2016}. In this case, each Weyl surface carries an Abelian second Chern number $C_2^A\in\mathbb{Z}$ of the U($1$) Berry curvature, which is equal to the sum of its linking number with all the other Weyl surfaces \cite{Lian2016}. Two natural questions are then whether the two 5d Weyl semimetal generalizations are related, and whether they play the role of intermediate phases during the TPT of certain gapped topological states of matter in 5d.

In this letter, we show the two 5d Weyl semimetal generalizations, namely, the Yang monopole and the linked Weyl surfaces in 5d, are closely related via the $\mathbf{TP}$ symmetry breaking, where $\mathbf{T}$ and $\mathbf{P}$ stand for time-reversal and space-inversion, respectively. We then demonstrate they also arise as intermediate phases in the TPT between 5d CI and NI, and between 5D TI and NI with particle-hole symmetry $\mathbf{C}$ that satisfies $\mathbf{C}^2=-1$ \cite{Qi2008,Kitaev2009,Ryu2010}. In analogy to 3d cases, the Weyl arcs on the boundary of the 5d Weyl semimetal \cite{Lian2016} naturally interpolate between the surface states of different gapped topological phases.

\section{Yang monopoles and linked Weyl surfaces}

In 3d, a Weyl semimetal is known as a semimetal which is gapless at several points in the BZ, i.e., Weyl points. The low energy bands near a Weyl point is generically given by a $2\times2$ Weyl fermion Hamiltonian $H_W(\mathbf{k})=\sum_{i=1}^3v_i(k_i-k^W_i)\sigma^i$ up to an identity term, where $\mathbf{k}$ is the momentum, and $\sigma^i$ ($i=1,2,3$) are the Pauli matrices. The Weyl point is located at $\mathbf{k}^W$, while the velocities $v_i\neq0$ ($i=1,2,3$) play the role of light speed. By defining the U($1$) Berry connection $a_i(\mathbf{k})=i\langle u_\mathbf{k}|\partial_{k_i}|u_\mathbf{k}\rangle$ of the valence (conduction) band wavefunction $|u_\mathbf{k}\rangle$, one can show the first Chern number of the Berry curvature $f_{ij}=\partial_{k_i}a_j-\partial_{k_j}a_i$ on a 2d sphere enclosing $\mathbf{k}^W$ is $C_1=\text{sgn}(v_1v_2v_3)=\pm1$, where $\text{sgn}(x)$ is the sign of $x$. Therefore, the Weyl point $\mathbf{k}^W$ can be viewed as a Dirac monopole of the Berry connection.

The first way of generalizing the Weyl semimetal to 5d is to replace the Weyl fermions above by the chiral Dirac fermions in 5d:
\begin{equation}\label{HY}
H_Y(\mathbf{k})=\sum_{i=1}^5v_i(k_i-k^Y_i)\gamma^i\ ,
\end{equation}
where $\mathbf{k}$ is now the 5d momentum, and $\gamma^i$ ($1\le i\le5$) are the $4\times4$ Gamma matrices satisfying the anticommutation relation $\{\gamma^i,\gamma^j\}=2\delta^{ij}$. The band structure of such a Hamiltonian is 4-fold degenerate at $\mathbf{k}^Y$, and is 2-fold degenerate everywhere else with a linear dispersion. The 2-fold degeneracy enables us to define a U($2$) Berry connection $a^{\alpha\beta}_i(\mathbf{k})=i\langle u^\alpha_\mathbf{k}|\partial_{k_i}|u^\beta_\mathbf{k}\rangle$, where $|u^\alpha_\mathbf{k}\rangle$ ($\alpha=1,2$) denote the two degenerate wavefunctions of the valence bands \cite{Wilczek1984}. One can then show the non-Abelian second Chern number $C_2^{NA}$ on a 4d sphere enclosing $\mathbf{k}^Y$ is
\begin{equation}\label{C2NA}
C_2^{NA}=\oint_{S^4}\frac{\mbox{d}^4\mathbf{k}\epsilon^{ijkl}[\mbox{tr}(f_{ij}f_{kl})-(\mbox{tr}f_{ij})(\mbox{tr}f_{kl})]}{32\pi^2}=\pm1\ ,
\end{equation}
where $f_{ij}=\partial_{k_i}a_j-\partial_{k_j}a_i-i[a_i,a_j]$ is the non-Abelian U($2$) Berry curvature. In this calculation, only the traceless SU(2) part of $f_{ij}$ contributes. Therefore, $\mathbf{k}^Y$ can be viewed as a Yang monopole in the BZ, which is the source of SU($2$) magnetic field in 5d \cite{Yang1978}.
However, the generic 2-fold degeneracy of Hamiltonian $H_Y(\mathbf{k})$ requires the system to have certain symmetries.
A common symmetry of this kind is the combined $\mathbf{TP}$ symmetry of time-reversal and inversion, which is anti-unitary and satisfies $(\mathbf{TP})^2=-1$ for fermions. Therefore, the Yang monopole 5d generalization is not in the same symmetry class as that of the generic 3d Weyl semimetal.

We remark here that the above 5d Yang monopole, together with the 3d Weyl point and the 2d Dirac point (e.g., in graphene), correspond exactly to the quaternion (pseudoreal), complex and real classes of the Wigner-Dyson threefold way \cite{Wigner1932,Dyson1962}, and the anti-unitary $\mathbf{TP}$ symmetry plays a key role in the classification. Basically, a matrix Hamiltonian $H(\mathbf{k})$ falls into these three classes if $(\mathbf{TP})^2=-1,0,+1$, respectively ($0$ stands for no $\mathbf{TP}$ symmetry), and one can show $d=5,3,2$ are the corresponding spacial dimensions where point-like gapless manifold in the BZ are stable. The minimal Hamiltonians of the three classes are listed in Tab. \ref{Threefold}. In particular, $(\mathbf{TP})^2=+1$ is possible for systems with a negligible spin-orbital coupling such as graphene, where the electrons can be regarded as spinless.

\begin{table}[bp]
  \centering
  \begin{tabular}{cccc}
  \hline
  $(\mathbf{TP})^2$ & \quad class & \quad$d$ & \qquad minimal model Hamiltonian \\
  \hline
  $+1$ &\quad$\mathbb{R}$ & \quad 2 & $H(\mathbf{k})=k_1\sigma^1+k_2\sigma^3$\\
  \hline
  $0$  &\quad$\mathbb{C}$ &\quad 3 & $\qquad H(\mathbf{k})=k_1\sigma^1+k_2\sigma^2+k_3\sigma^3$\\
  \hline
  $-1$  &\quad $\mathbb{Q}$ &\quad 5 & $H(\mathbf{k})=\sum_{i=1}^5 k_i\gamma^i$\\
  \hline
  \end{tabular}
  \caption{A Hamiltonian $H(\mathbf{k})$ with $(\mathbf{TP})^2=+1,0,-1$ is in the real ($\mathbb{R}$), complex ($\mathbb{C}$) and quaternion ($\mathbb{Q}$) classes of the Wigner-Dyson three-fold way, respectively, and the Hamiltonians of 2d Dirac point, 3d Weyl point and 5d Yang monopole shown here exactly fall into these three classes.}\label{Threefold}
\end{table}

The second 5d Weyl semimetal generalization requires no symmetry (other than the translational symmetry), thus is in the same symmetry class with the 3d Weyl semimetal. Its band structure is non-degenerate except for a few closed submanifolds $\mathcal{M}_j$ called Weyl surfaces where two bands cross each other\cite{Lian2016}. The effective Hamiltonian near each $\mathcal{M}_j$ involves only the two crossing bands and takes the $2\times2$ form $H_{W}(\mathbf{k})=\xi_0(\mathbf{k})+\sum_{i=1}^3\xi_i(\mathbf{k})\sigma^i$. Therefore, $\mathcal{M}_j$ is locally determined by 3 conditions $\xi_i(\mathbf{k})=0$ ($i=1,2,3$). In one band $\alpha$ of the two associated with $\mathcal{M}_j$, one can define a U($1$) Berry connection $a^{(\alpha)}_i(\mathbf{k})=i\langle u^\alpha_\mathbf{k}|\partial_{k_i}|u^\alpha_\mathbf{k}\rangle$ with its wavefunction $|u^\alpha_\mathbf{k}\rangle$, and define the U($1$) second Chern number of $\mathcal{M}_j$ in band $\alpha$ on a 4d closed manifold $\mathcal{V}$ that only encloses $\mathcal{M}_j$ as
\begin{equation}
C_2^A(\mathcal{M}_j,\alpha)=\oint_\mathcal{V}\frac{\mbox{d}^4\mathbf{k}\epsilon^{ijkl}f^{(\alpha)}_{ij}f^{(\alpha)}_{kl}}{32\pi^2}\in\mathbb{Z}\ ,
\end{equation}
where $f^{(\alpha)}_{ij}$ is the Berry curvature of $a^{(\alpha)}_{i}$. Remarkably, we showed in an earlier paper that \cite{Lian2016}
\begin{equation}\label{Link}
C_2^A(\mathcal{M}_j,\alpha)=\sum_{\ell\in\alpha,\ell\neq j}\Phi(\mathcal{M}_j,\mathcal{M}_\ell)\ ,
\end{equation}
where $\Phi(\mathcal{M}_j,\mathcal{M}_\ell)$ is the linking number between $\mathcal{M}_j$ and $\mathcal{M}_\ell$ in the 5d BZ, and $\mathcal{M}_\ell$ runs over all the Weyl surfaces associated with band $\alpha$.

\begin{figure}[tbp]
\begin{center}
\includegraphics[width=3.2in]{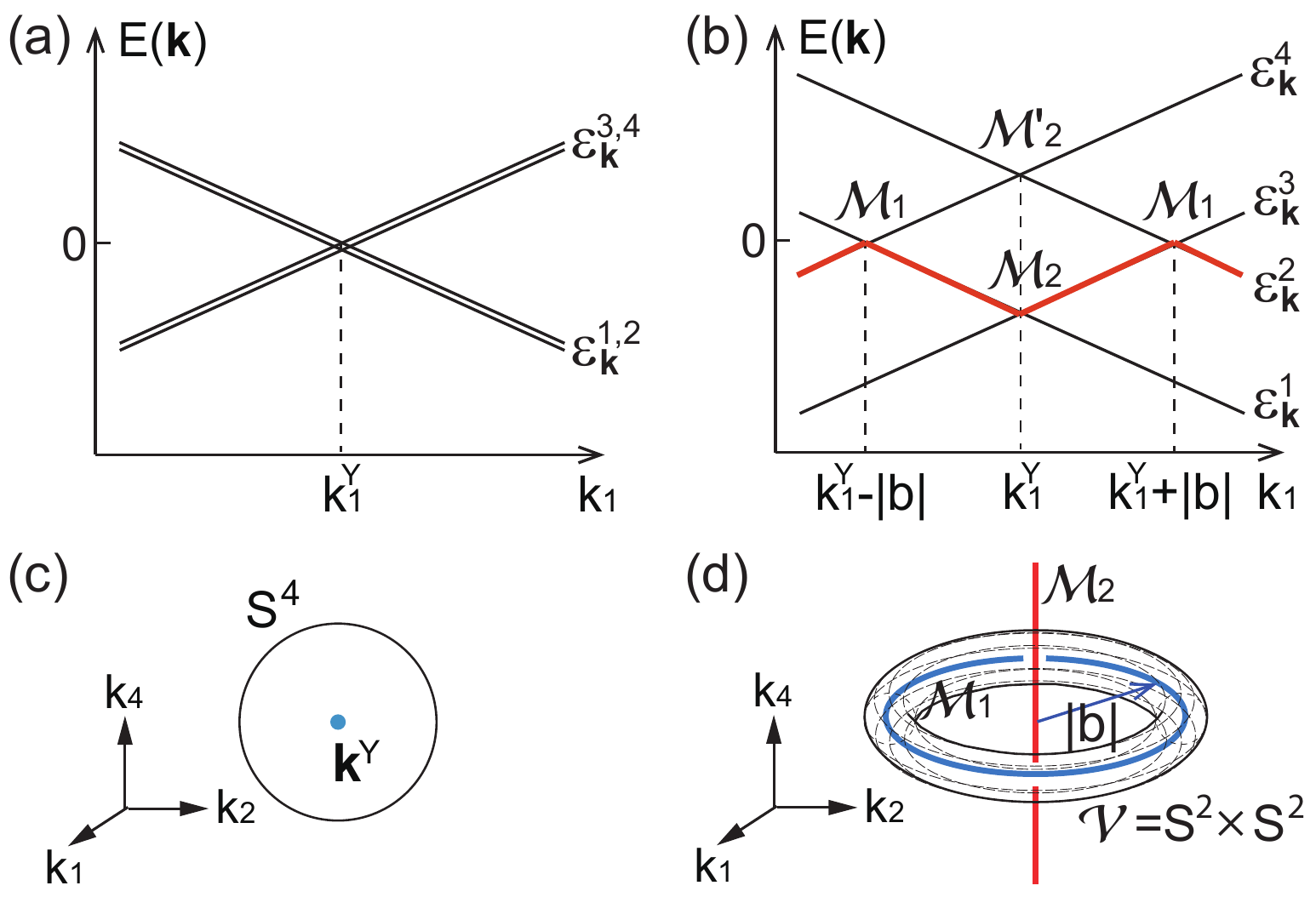}
\end{center}
\caption{(color online) (a) Doubly degenerate band structure near a Yang monopole plotted at $\tilde{k}_{i}=0$ ($2\le i\le5$). (b) The band structure and Weyl surfaces in the presence of the $\mathbf{TP}$ breaking $b$ term. (c) Yang monopole enclosed by $S^4$ shown in the 3d subspace $\tilde{k}_3=\tilde{k}_5=0$. (d) The hopf link of Weyl surfaces $\mathcal{M}_1$ and $\mathcal{M}_2$ shown in 3d subspace $\tilde{k}_3=\tilde{k}_5=0$ (thick red and blue lines), and a 4d manifold $\mathcal{V}=S^2\times S^2$ enclosing Weyl surface $\mathcal{M}_1$ (appearing as a torus). }
\label{Yang}
\end{figure}

The relation between the above two 5d generalizations can be most easily seen in the following $4$-band model with Hamiltonian
\begin{equation}\label{HYp}
H_{Y}'(\mathbf{k})=\sum_{i=1}^5(k_i-k^Y_i)\gamma^i+b\frac{i[\gamma^4,\gamma^5]}{2}\ ,
\end{equation}
where $b$ is a real parameter that breaks the $\mathbf{TP}$ symmetry. When $b=0$, the Hamiltonian reduces to the Yang monopole Hamiltonian $H_Y(\mathbf{k})$ in Eq. (\ref{HY}), where we have set all the velocities to $v_i=1$. When $b\neq0$, the $\mathbf{TP}$ symmetry is broken, and the Yang monopole necessarily evolves into linked Weyl surfaces. This can be seen explicitly by deriving the energy spectrum $\epsilon^\alpha_\mathbf{k}=\pm[((\tilde{k}_1^2+\tilde{k}_2^2+\tilde{k}_3^2)^{1/2}\pm b)^2+\tilde{k}_4^2+\tilde{k}_5^2]^{1/2}$, where we have defined $\tilde{k}_i=k_i-k_i^Y$ ($1\le i\le5$). Here $1\le\alpha\le4$ denotes the $\alpha$-th band in energies.
Fig. \ref{Yang}(a) and Fig. \ref{Yang}(b) show the band structures for $b=0$ and $b\neq0$, respectively, where $\tilde{k}_2,\tilde{k}_3,\tilde{k}_4,\tilde{k}_5$ are assumed zero. In the $b\neq0$ case, one can readily identify three Weyl surfaces: $\mathcal{M}_1$ between bands $\epsilon^2_\mathbf{k}$ and $\epsilon^3_\mathbf{k}$, $\mathcal{M}_2$ between bands $\epsilon^1_\mathbf{k}$ and $\epsilon^2_\mathbf{k}$ and $\mathcal{M}_2'$ between bands $\epsilon^3_\mathbf{k}$ and $\epsilon^4_\mathbf{k}$ (see Fig. \ref{Yang}(b)). $\mathcal{M}_1$ is a 2d sphere given by $\tilde{k}_1^2+\tilde{k}_2^2+\tilde{k}_3^2=b^2$ and $\tilde{k}_4=\tilde{k}_5=0$, while $\mathcal{M}_2$ and $\mathcal{M}_2'$ coincide and are a 2d plane given by $\tilde{k}_1=\tilde{k}_2=\tilde{k}_3=0$. In particular, the second band $\epsilon_\mathbf{k}^2$ (thick red line in Fig. \ref{Yang}(b)) is associated with $\mathcal{M}_1$ and $\mathcal{M}_2$, which form a Hopf link in 5d as can be seen in the 3d subspace $k_3=k_5=0$ plotted in Fig. \ref{Yang}(d). In the limit $b\rightarrow0$, the radius of $\mathcal{M}_1$ contracts to zero, so $\mathcal{M}_1$ collapses onto $\mathcal{M}_2$ (and $\mathcal{M}_2'$) and becomes the 4-fold degenerate Yang monopole in Fig. \ref{Yang}(c).
One can add other small $\mathbf{TP}$ breaking terms to Eq. (\ref{HYp}), and the above picture remains topologically unchanged.

\begin{figure}[tbp]
\begin{center}
\includegraphics[width=3.2in]{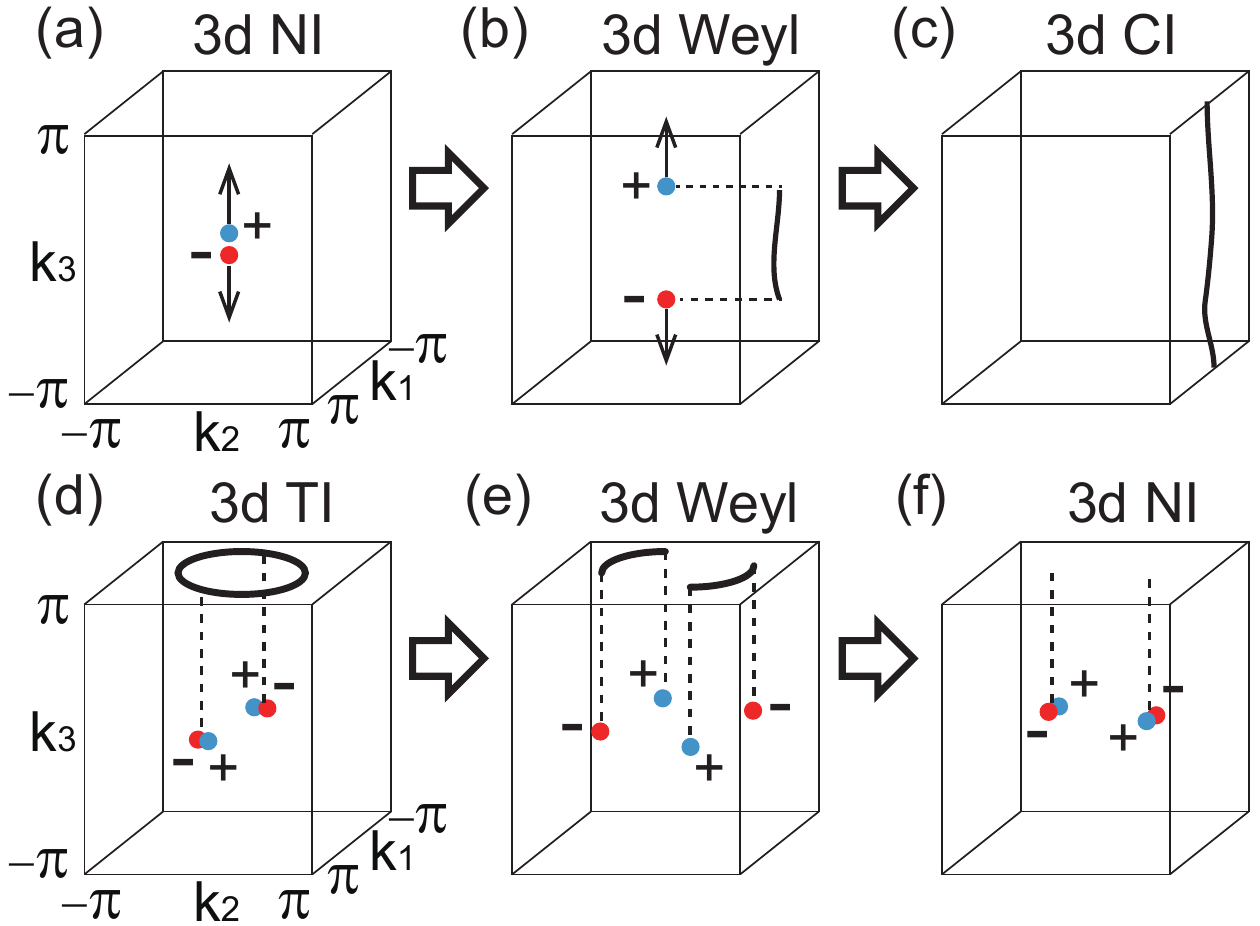}
\end{center}
\caption{(color online) (a)-(c) TPT from a 3d NI to a 3d CI via creation and annihilation of a pair of Weyl points in the BZ. (d)-(f) TPT from a 3d TI to a 3d NI, which involves winding of (multiples of) four Weyl points.}
\label{3DWeyl}
\end{figure}

Due to the $\mathbf{TP}$ symmetry breaking, the U(2) gauge field $a_i(\mathbf{k})$ is broken down to two U(1) gauge fields $a_i^{(1)}(\mathbf{k})$ and $a_i^{(2)}(\mathbf{k})$ in bands $\epsilon_\mathbf{k}^1$ and $\epsilon_\mathbf{k}^2$. One can easily check the Abelian second Chern number of $\mathcal{M}_1$ calculated from $a_i^{(2)}(\mathbf{k})$ is $C_2^A(\mathcal{M}_1,2)=1$, which is defined on 4d manifold $\mathcal{V}$ with topology $S^2\times S^2$ as shown in Fig. \ref{Yang}(d) \cite{Lian2016}. This is closely related to the non-Abelian second Chern number $C_2^{NA}=1$ of the Yang monopole before symmetry breaking. In fact, ignoring the gauge invariance, we can still define the U(2) gauge field $a_i^{\alpha\beta}(\mathbf{k})$ using the two valence bands of Hamiltonian $H_{Y}'(\mathbf{k})$, which is singular on $\mathcal{M}_1$ but not on $\mathcal{M}_2$ (since $\mathcal{M}_2$ is between the two bands defining the U(2) Berry connection), and still satisfies $C_2^{NA}=1$ on a sphere $S^4$ enclosing $\mathcal{M}_1$. The sphere $S^4$ can be deformed adiabatically into $\mathcal{V}$ in Fig. \ref{Yang}(d), so we also have $C_2^{NA}=1$ on $\mathcal{V}$.
To see $C_2^{NA}$ is equal to $C_2^A(\mathcal{M}_1,2)$, we can take the limit $\mathcal{V}$ is a thin "torus" $S^2\times S^2$, i.e., its smaller radius (distance to $\mathcal{M}_1$) tends to zero. In this limit, one will find $\int_\mathcal{V}\mbox{d}^4\mathbf{k}\epsilon^{ijkl}f^{12}_{ij}f^{21}_{kl}=0$, namely, the off-diagonal elements of field strength $f_{ij}$ do not contribute (see Appendix \ref{A1}). So $C_{2}^{NA}$ is solely given by the diagonal field strengths $f^{11}_{ij}$ and $f^{22}_{ij}$, which can be roughly identified with U(1) Berry curvatures of band $1$ and $2$. By calculations, one can show $\epsilon^{ijkl}f^{11}_{ij}f^{11}_{kl}=\epsilon^{ijkl}\mbox{tr}f_{ij}\mbox{tr}f_{kl}=0$. A heuristic understanding of this is the Berry curvature $f^{11}_{ij}$ of band $1$ sees only $\mathcal{M}_2$, while the U(1) trace Berry curvature $\mbox{tr}f_{ij}$ sees only $\mathcal{M}_1$, so both of them do not see linked Weyl surfaces and have zero contribution to the second Chern number. One can then readily show
$C_2^{NA}=\int_\mathcal{V}{\mbox{d}^4\mathbf{k}\epsilon^{ijkl}f^{22}_{ij}f^{22}_{kl}}/{32\pi^2} =\int_\mathcal{V}{\mbox{d}^4\mathbf{k}\epsilon^{ijkl}f^{(2)}_{ij}f^{(2)}_{kl}}/{32\pi^2}=C_2^A(2,\mathcal{M}_1)$. We note that in this limit where $\mathcal{V}$ is closely attached to $\mathcal{M}_1$, only the diagonal elements of $f_{ij}$ contributes, while in the Yang monopole case which is spherically symmetric, the diagonal and off-diagonal elements are equally important \cite{Yang1978}.

In high energy physics, a U(2) gauge symmetry can be spontaneously broken down to U(1)$\times$U(1) via the Georgi-Glashow mechanism \cite{Georgi1972} with an isospin $1$ Higgs field. In 5d space, SU(2) gauge fields are associated with point-like Yang monopoles, while U(1) gauge fields are associated with monopole 2-branes (codimension 3 objects). We conjecture that a gauge symmetry breaking from U(2) to U(1)$\times$U(1) in 5d will always break an SU(2) Yang monopole into two linked U(1) monopole $2$-branes $\mathcal{M}_1$ and $\mathcal{M}_2$, where $\mathcal{M}_1$ is coupled to one of the two U(1) gauge fields, while $\mathcal{M}_2$ is coupled to both U(1) gauge fields with opposite monopole charges.

\section{Topological Phase Transitions in 5d}

It is known that 3d Weyl semimetals play an important role in 3d TPTs. An example is the TPT of 3d Chern insulator (CI) with no symmetry, which is characterized by three integers $(n_1,n_2,n_3)$, with $n_i$ being the first Chern number in the plane orthogonal to $k_i$ in the BZ \cite{Kohmoto1992,Haldane2004}. The CI becomes a normal insulator (NI) when all $n_i=0$. The TPT from a 3d NI to a $(0,0,1)$ CI involves an intermediate Weyl semimetal phase as shown in Fig. \ref{3DWeyl}(a)-(c). By creating a pair of Weyl points with opposite monopole charges and annihilating them after winding along a closed cycle in $k_3$ direction, one creates a Berry flux quanta in the $k_1$-$k_2$ plane, and $n_3$ increases by one \cite{Haldane2004}. At the same time, a fermi arc arises on the real space boundary connecting the projections of the two Weyl points \cite{Wan2011}, which finally becomes a closed fermi loop along $k_3$.
Another example is the TPT from TI to NI, which are the two phases in the $\mathbb{Z}_2$ classification of 3d TRI insulators \cite{Fu2007,Qi2008}. When the inversion symmetry is broken, an intermediate TRI Weyl semimetal arises \cite{Murakami2007,Murakami2016}, which contains (multiples of) 4 Weyl points as shown in Fig. \ref{3DWeyl}(d)-(f). The TPT is done by creating two pairs of Weyl points with opposite charges, winding them along a loop that encloses a TRI point (e.g., $\Gamma$ point), then annihilating them in pairs with their partners exchanged. Meanwhile, the fermi surface loop of the Dirac surface states of TI breaks into two fermi arcs connecting the 4 Weyl points, which vanish when all the Weyl points are gone.

\begin{figure}[tbp]
\begin{center}
\includegraphics[width=3.2in]{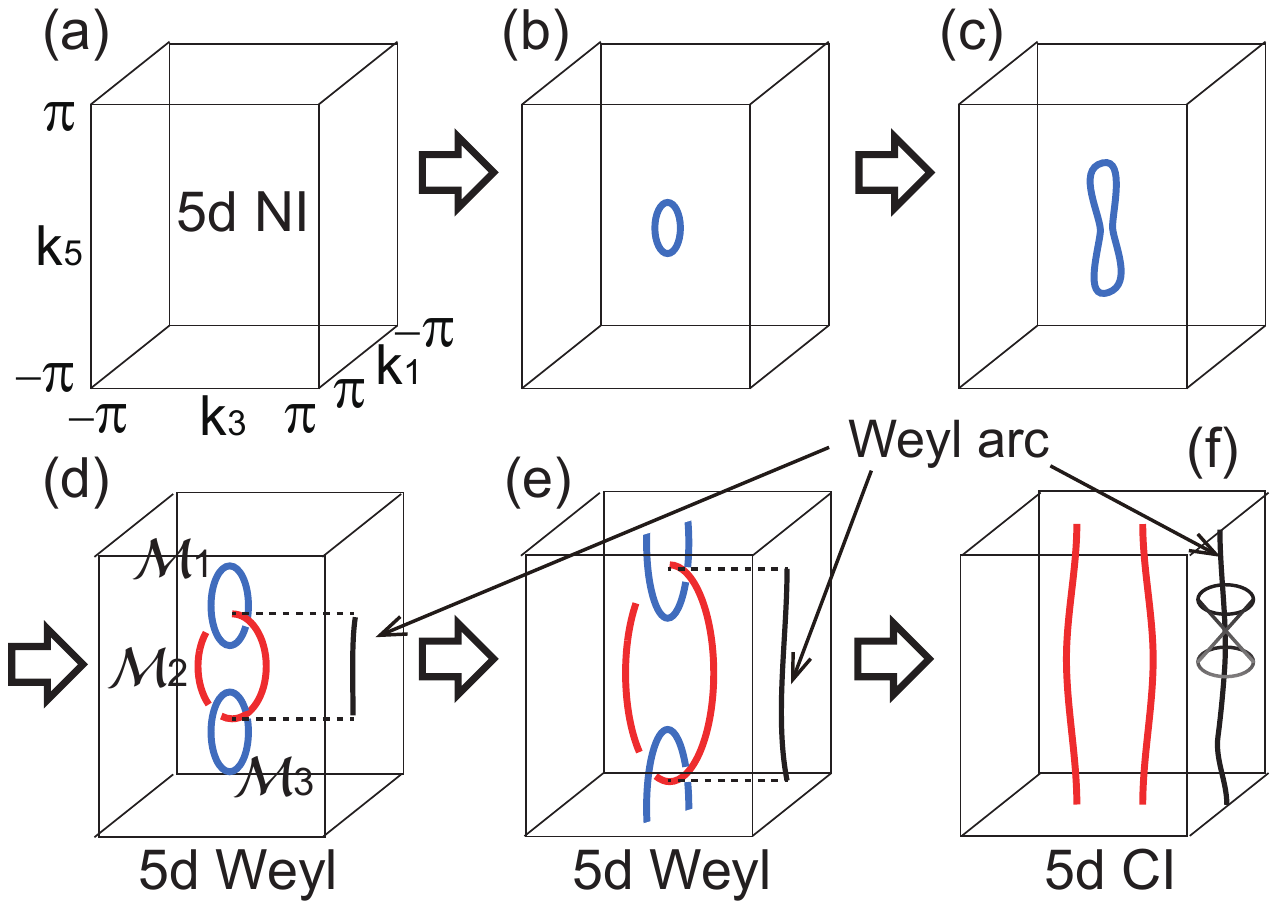}
\end{center}
\caption{(color online) The evolution of Weyl surfaces in the 5d TPT from NI to CI, plotted in 3d subspace $k_2=k_4=0$. The blue loops are Weyl surfaces between bands $2$ and $3$, while the red loop is that between bands $1$ and $2$.}
\label{QAHI}
\end{figure}

Similarly, the 5d TPTs involve creation of 5d Weyl semimetal phases. We first examine the TPT of 5d CIs with no symmetry, which are characterized by 5 second Chern numbers $n_i$ in the 4d hyperplanes of the BZ orthogonal to $k_i$ ($1\le i\le5$), and 10 first Chern numbers $n_{ij}$ in the 2d planes parallel to $k_i$ and $k_j$ ($1\le i<j\le5$). The second Chern numbers $n_i$ are even under both $\mathbf{T}$ and $\mathbf{P}$ transformations, while the first Chern numbers $n_{ij}$ are odd under $\mathbf{T}$ and even under $\mathbf{P}$. Here we shall show that changes of the five $n_i$ will involve creation and annihilation of linked Weyl surfaces in the BZ. A simple example without $\mathbf{TP}$ symmetry is the following 4-band Hamiltonian
\begin{equation}
H_{QH}(\mathbf{k})=\sum_{i=1}^5\xi_i(\mathbf{k})\gamma^i+b\frac{i[\gamma^3,\gamma^4]}{2}\ ,
\end{equation}
where $\xi_i(\mathbf{k})=\sin k_i$ for $1\le i\le 4$, and $\xi_5(\mathbf{k})=m+\sum_{i=1}^4(1-\cos k_i)+\eta(1-\cos k_5)$. Here $m$ is a tuning parameter, while $0\le b<\eta<1-b$.
We shall label each band by its order in energies, and assume the lower two bands are occupied. Through an analysis similar to we did below Eq. (\ref{HYp}), the Weyl surfaces between bands $2$ and $3$ are given by $\xi_1^2+\xi_2^2+\xi_5^2=b^2$ and $\xi_3=\xi_4=0$, while those between bands $1$ and $2$ (also $3$ and $4$) are given by $\xi_1=\xi_2=\xi_5=0$. These two kinds of Weyl surfaces are drawn as blue and red in the 3d subspace $k_2=k_4=0$ of the BZ shown in Fig. \ref{QAHI}, respectively, where they appear as 1d loops.

When $m>b$, the system is a 5d NI with all $n_i$ and $n_{ij}$ zero and no Weyl surfaces (Fig. \ref{QAHI}(a)). The TPT to a 5d CI with $n_5=1$ is driven by decreasing $m$, which experiences the following stages: when $-b<m<b$, a Weyl surface between bands $2$ and $3$ arise around the origin, which is topologically a 2d sphere in the $k_3=k_4=0$ hyperplane (Fig. \ref{QAHI}(b), (c)). When $b-2\eta<m<-b$, as shown in Fig. \ref{QAHI}(d), the 2d sphere between band $2$ and $3$ splits into two smaller spheres $\mathcal{M}_1$ and $\mathcal{M}_3$ (blue) in the $k_3=k_4=0$ hyperplane, while another 2d sphere Weyl surface $\mathcal{M}_2$ (red) between bands $1$ and $2$ is created in $k_1=k_2=0$ plane, which is linked to both $\mathcal{M}_1$ and $\mathcal{M}_3$. As $m$ is further decreased, $\mathcal{M}_1$ and $\mathcal{M}_3$ will move along $\pm k_5$, respectively, and finally merge into a single Weyl surface when $-b-2\eta<m<b-2\eta$ (Fig. \ref{QAHI}(e)). This Weyl surface then shrinks to zero, and the system becomes a 5d CI with $n_5=1$ for $b-2<m<-b-2\eta$, leaving a cylindrical Weyl surface $\mathcal{M}_2$ between bands $1$ and $2$ (also one between bands $3$ and $4$, Fig. \ref{QAHI}(f)).
We note that if $b=0$, the $\mathbf{TP}$ symmetry is restored, and the two blue Weyl surfaces $\mathcal{M}_1$ and $\mathcal{M}_3$ will collapse into two Yang monopoles of opposite monopole charges $C_2^{NA}$. The TPT process then becomes the creation, winding and annihilation of two Yang monopoles.


This TPT is also accompanied with a surface state evolution from trivial to nontrivial. It has been shown \cite{Lian2016} that a 5d Weyl semimetal with linked Weyl surfaces contain protected Weyl arcs in the 4d momentum space of surface states, which have linear dispersions in the other $3$ directions perpendicular to the arc. By taking an open boundary condition along $k_3$ direction, one can obtain a Weyl arc on the 4d boundary connecting the projections of the two Weyl surface hopf links (Fig. \ref{QAHI}(d), (e)). When the system becomes a CI, the Weyl arc develops into a non-contractible Weyl loop along $k_5$ as expected.

The second example is the TPT between $\mathbf{TP}$ breaking 5d insulators with particle-hole symmetry $\mathbf{C}$ satisfying $\mathbf{C}^2=-1$, which are shown to be classified by $\mathbb{Z}_2$ into 5d TIs and NIs \cite{Qi2008,Kitaev2009,Ryu2010}. Here we consider an eight-band model Hamiltonian of a 5d TI as follows:
\begin{equation}
H_{TI}(\mathbf{k})
=\sum_{i=1}^6\zeta_i(\mathbf{k})\Gamma^i+H_A\ ,
\end{equation}
where $\Gamma^i$ ($1\le i\le7$) are the $8\times8$ Gamma matrices so chosen that $\Gamma^{1}$, $\Gamma^{2}$, $\Gamma^{3}$, $\Gamma^{7}$ are real and $\Gamma^{4}$, $\Gamma^{5}$, $\Gamma^{6}$ are imaginary, $\zeta_i(\mathbf{k})=\sin k_i$ for $1\le i\le5$, $\zeta_6(\mathbf{k})=m+\sum_{i=1}^5t_i(1-\cos k_i)$ with $t_i>0$, and
\begin{equation}
H_A=i\eta_0\Gamma^{1}\Gamma^2\Gamma^7+\eta_1\Gamma^7\sin k_5+i\eta_2\Gamma^{3}\Gamma^4\Gamma^5+i\eta_3\Gamma^{3}\Gamma^4
\end{equation}
is a symmetry breaking perturbation. The
$\mathbf{T}$, $\mathbf{P}$ and $\mathbf{C}$ transformation matrices are given by $\mathcal{T}=\Gamma^4\Gamma^5\Gamma^7$, $\mathcal{P}=i\Gamma^6$ and $\mathcal{C}=\Gamma^4\Gamma^5$, and a Hamiltonian $H(\mathbf{k})$ will have these symmetries if $\mathcal{T}^\dag H(-\mathbf{k})\mathcal{T}=H^*(\mathbf{k})$, $\mathcal{P}^\dag H(-\mathbf{k})\mathcal{P}=H(\mathbf{k})$ and $\mathcal{C}^\dag H(-\mathbf{k})\mathcal{C}=-H^*(\mathbf{k})$, respectively. It is then easy to see that $H_A$ respects the $\mathbf{C}$ symmetry but breaks $\mathbf{T}$, $\mathbf{P}$ and $\mathbf{TP}$ symmetries. In particular, only $\eta_1$ and $\eta_2$ breaks the $\mathbf{TP}$ symmetry.

\begin{figure}[t]
\begin{center}
\includegraphics[width=3.2in]{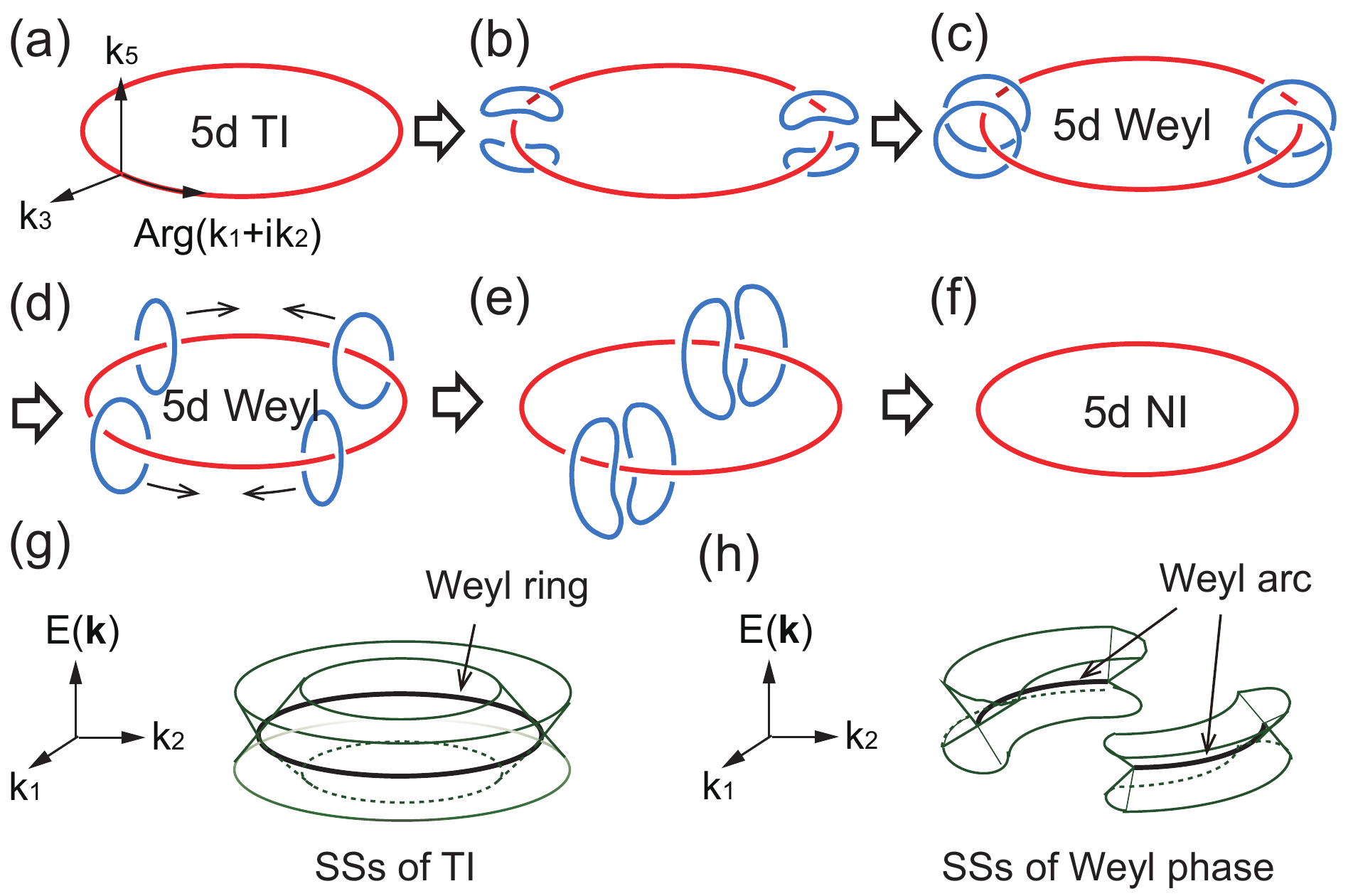}
\end{center}
\caption{(color online) (a)-(f) The TPT from 5d TI to NI shown in 3d subspace $k_1^2+k_2^2=\eta_0^2$ and $k_4=0$. The small blue loops are the Weyl surfaces at the Fermi energy between bands $4$ and $5$, while the large red loop is that between bands $3$ and $4$. (g)-(h) The evolution of surface states from a Weyl ring to two Weyl arcs during the TPT.}
\label{TINI}
\end{figure}

In the absence of $H_A$, the system is a 5d NI if $m>0$, and is a 5D TI if $m<0$. With the symmetry breaking term $H_A$, the calculation of the band structure of $H_{TI}(\mathbf{k})$ becomes more complicated.
For simplicity, we shall only examine the limiting case where $(t_1/t_2)^2+\eta_0^2<1$, $|\eta_0|\gg|\eta_1|$, $|\eta_0|\gg|\eta_2|$ and $|\eta_0\eta_1|\gg|\eta_3|$ (with $t_1$ and $t_2$ as defined in the expression of $\zeta_6(\mathbf{k})$). We shall label each band by its order in energies, and keep the Fermi energy at zero, i.e., between band $4$ and $5$, as required by the $\mathbf{C}$ symmetry. To a good approximation, the Weyl surfaces between bands $4$ and $5$ are given by $\zeta_1^2+\zeta_2^2=\eta_0^2$, $\zeta_3^2+\zeta_4^2+\zeta_5^2=\eta_2^2$ and $\zeta_6=0$, while those between bands $5$ and $6$ (also between $3$ and $4$) are given by $\zeta_3=\zeta_4=\zeta_5=0$.
The TPT can be driven by tuning $m$ from negative (TI) to positive (NI), and the evolution of these low energy Weyl surfaces are illustrated in Fig. \ref{TINI} (a)-(f) in the 3d subspace $k_1^2+k_2^2=\eta_0^2$ and $k_4=0$. The small blue loops are the images of Weyl surfaces between bands $4$ and $5$, while the red loop at $k_3=k_5=0$ is that between bands $5$ and $6$. At first, two pairs of blue Weyl surfaces arise unlinked (panel (b)). As $m$ increases, they merge into four new Weyl surfaces linked with the red Weyl surface, which then wind around the red Weyl surface and merge into unlinked pairs again with their partners exchanged (panels (c)-(e)). Finally, the four unlinked blue Weyl surfaces contract to zero, and the system becomes a 5d NI.
Similar to the CI case, if $\eta_1=\eta_2=0$, the $\mathbf{TP}$ symmetry is recovered, and the 4 blue Weyl surfaces will collapse into 4 Yang monopoles. The TPT process then involves winding of Yang monopoles instead of linked Weyl surfaces. The topological surface states of the system also involves a topological transition during the TPT. The topological surface states of a noncentrosymmetric 5d TI is generically a "Weyl ring" as shown in Fig.\ref{TINI}(g). The TPT then breaks into two Weyl arcs (panel (h)), which finally vanish when entering the NI phase.

In conclusion, we show that 5d Weyl semimetals with Yang monopoles are protected by the $\mathbf{TP}$ symmetry, and generically reduce to 5d Weyl semimetals with linked Weyl surfaces in the presence of $\mathbf{TP}$ symmetry breaking. We therefore expect that Yang monopoles generically break into linked U(1) monopole 2-branes in 5d theories of gauge symmetry breaking from U(2) to U(1)$\times$U(1). As gapless states carrying the second Chern number, they emerge as intermediate phases in the TPTs between CIs and NIs or between TIs and NIs in 5d space, generalizing the connection between gapless and gapped topological phases in 3d \cite{Murakami2007,Murakami2016}.

\begin{acknowledgments}
\emph{Acknowledgments}. We would like to thank Jing-Yuan Chen for helpful discussions during the research. This work is supported by the NSF under grant number DMR-1305677.
\end{acknowledgments}

%

\appendix

\section{Derivation of $C_2^{NA}=C_2^A(\mathcal{M}_1,2)$ on manifold $\mathcal{V}$}\label{A1}
It is sufficient to do the calculation in the limit $\mathcal{V}$ is thin, i.e., close to Weyl surface $\mathcal{M}_1$. For the model given in Eq. (\ref{HYp}), such a 4d manifold $\mathcal{V}$ can be given by $(\kappa-b)^2+\tilde{k}_4^2+\tilde{k}_5^2=\epsilon^2$, where $\kappa=\sqrt{\tilde{k}_1^2+\tilde{k}_2^2+\tilde{k}_3^2}$, and $\epsilon\ll b$. Using Gamma matrices defined in \cite{Lian2016}, the wave function $|u^2_\mathbf{k}\rangle$ is given by
\begin{equation}
\begin{split}
&|u^2_\mathbf{k}\rangle=(\cos\frac{\theta}{2}\cos\frac{\alpha}{2},\sin\frac{\theta}{2}\cos\frac{\alpha}{2}e^{i\phi}, \\ &\qquad\cos\frac{\theta}{2}\sin\frac{\alpha}{2}e^{i\psi},\sin\frac{\theta}{2}\cos\frac{\alpha}{2}e^{i\phi+i\psi})\ ,
\end{split}
\end{equation}
while $|u^1_\mathbf{k}\rangle$ is well-approximated by
\begin{equation}
|u^1_\mathbf{k}\rangle=(\sin\frac{\theta}{2},-\cos\frac{\theta}{2}e^{i\phi},0,0)\ ,
\end{equation}
where we have defined the angles $\alpha, \psi, \theta,\phi$ by $\sin\alpha e^{i\psi}=(k_4+ik_5)/\epsilon$, and $\sin \theta e^{i\phi}=(k_1+ik_2)/\kappa$. This approximation basically ignores the dependence of $|u^1_\mathbf{k}\rangle$ on $k_4$ and $k_5$, which is valid since $|\tilde{k}_{4,5}|<\epsilon\ll b$, and $|u^1_\mathbf{k}\rangle$ is nonsingular at $\mathcal{M}_1$. The nonzero components of U(2) Berry connection $a^{\alpha\beta}_\mathbf{k}$ can then be shown to be
\begin{equation}
\begin{split}
&a^{11}_\phi=-\frac{1+\cos\theta}{2},a^{22}_\phi=-\frac{1-\cos\theta}{2}, \ a^{22}_\psi=-\frac{1-\cos\alpha}{2},\\
&a^{21}_\theta=\frac{i}{2}\cos\theta\cos\frac{\alpha}{2}\ ,\ a^{21}_\phi=\frac{1}{2}\sin\theta\cos\frac{\alpha}{2}\ ,
\end{split}
\end{equation}
with $a^{12}_i=a^{21*}_i$. It is then straightforward to calculate the non-Abelian field strengths $f^{\alpha\beta}_{ij}$. In particular, one can prove that $\epsilon^{ijkl}f^{12}_{ij}f^{21}_{kl}=[\sin2\theta(1-\cos\alpha)\sin\alpha]/8$, which gives $0$ when integrated over the four angles. Therefore, the off-diagonal components of $f_{ij}$ have no contribution to the second Chern number $C_2^{NA}$. Further, one can show $f_{\theta\phi}^{22}=-f_{\theta\phi}^{11}=f_{\theta\phi}^{(2)}+(\sin 2\theta\cos\alpha)/4$ and $f_{\alpha\psi}^{22}=f_{\alpha\psi}^{(2)}=(\sin\alpha)/2$ are the only rest nonzero terms, where $f_{ij}^{(1)}$ and $f_{ij}^{(2)}$ are the U(1) Berry connection in band $1$ and $2$, respectively. Therefore, we have $\epsilon^{ijkl}f^{11}_{ij}f^{11}_{kl}=\epsilon^{ijkl}\mbox{tr}f_{ij}\mbox{tr}f_{kl}=0$, and $\epsilon^{ijkl}f^{22}_{ij}f^{22}_{kl}=\epsilon^{ijkl}f^{(2)}_{ij}f^{(2)}_{kl}+\sin 2\theta\sin2\alpha$. The non-Abelian second Chern number is then
\begin{equation}
\begin{split}
C_2^{NA}&=\int_0^\pi\mbox{d}\theta\int_0^{2\pi} \mbox{d}\phi\int_0^\pi \mbox{d}\alpha\int_0^{2\pi}\mbox{d}\psi\frac{\epsilon^{ijkl}f^{22}_{ij}f^{22}_{kl}}{32\pi^2} \\ &=\oint_{\mathcal{V}}\frac{\mbox{d}^4\mathbf{k}\epsilon^{ijkl}f^{(2)}_{ij}f^{(2)}_{kl}}{32\pi^2}=C_2^A(2,\mathcal{M}_1)=1\ .
\end{split}
\end{equation}
If we rewrite the non-Abelian field strength as $f_{ij}=f_{ij}^at^a$ where $t^a=(1,\sigma^1,\sigma^2,\sigma^3)/2$ ($a=0,1,2,3)$ are the generator of U(2), the non-Abelian second Chern number on $\mathcal{V}$ can be expressed as
\[C_2^{NA}=-c^0+c^1+c^2+c^3 ,\]
where we have defined $c^a=\int_\mathcal{V} \mbox{d}^4\mathbf{k}\epsilon^{ijkl}f^{a}_{ij}f^{a}_{kl}/64\pi^2$. In the limit $\mathcal{V}$ is close to $\mathcal{M}_1$, the above calculations tell us that $c_0=c_1=c_2=0$, and $c_3=C_2(2,\mathcal{M}_1)=1$. In contrast, in the Yang monopole case which is rotationally symmetric, one can show $c_0=0$, and $c_1=c_2=c_3=1/3$. Therefore, the $TP$ symmetry breaking also breaks the symmetry between $c_1$, $c_2$ and $c_3$.

\section{Weyl surfaces of model Hamiltonian Eq. (7)}
Compared to the four-band model in Eq. (6) which can be easily diagonalized, the eight-band model $H_{TI}(\mathbf{k})$ in Eq. (7) has a band structure more difficult to calculate. Here we present an easier way to examine the band structure with the assumptions $(t_1/t_2)^2+\eta_0^2<1$, $|\eta_0|\gg|\eta_1|$, $|\eta_0|\gg|\eta_2|$ and $|\eta_0\eta_1|\gg|\eta_3|$.

For the moment we shall assume $\eta_3=0$. To solve the Schr\"{o}dinger equation $H_{TI}|\psi\rangle=E|\psi\rangle$, one can first rewrite it into $H_{TI}^2|\psi\rangle=E^2|\psi\rangle$, which reduces to
\begin{equation}
\Big(E^2-\sum_{i=1}^6\xi_i^2-\eta_0^2-\eta_1^2\sin^2 k_5-\eta_2^2\Big)|\psi\rangle=\left(\Lambda_0+\Lambda_2\right)|\psi\rangle
\end{equation}
after making use of the properties of Gamma matrices, where we have defined
\[\Lambda_0=2\eta_0\left(\zeta_1\Gamma^2\Gamma^7-\zeta_2\Gamma^1\Gamma^7+\eta_1\sin k_5\Gamma^1\Gamma^2\right)\ ,\]
\[\Lambda_2=2\eta_2\left(\zeta_3\Gamma^4\Gamma^5-\zeta_4\Gamma^3\Gamma^5+\zeta_5\Gamma^3\Gamma^4\right)\ .\]
One can easily show that $\Lambda_0^2=4\eta_0^2(\zeta_1^2+\zeta_2^2+\eta_1^2\sin^2 k_5)=\eta_0^2\chi_0^2$, $\Lambda_2^2=4\eta_2^2(\zeta_3^2+\zeta_4^2+\zeta_5^2)=\eta_2^2\chi_2^2$, and $[\Lambda_0,\Lambda_2]=0$. Therefore, they can be simultaneously diagonalized, i.e., $\Lambda_0=\pm\eta_0\chi_0$, $\Lambda_2=\pm\eta_2\chi_2$. One then obtain the energy spectrum of the eight bands as
\begin{equation}
E=\pm\sqrt{\left(\eta_0\pm\chi_0\right)^2+\left(\eta_2\pm\chi_2\right)^2+\zeta_6^2}\ .
\end{equation}
One can then see the Weyl surfaces between bands 4 and 5 are given by $\chi_0^2=\zeta_1^2+\zeta_2^2+\eta_1^2\sin^2 k_5=\eta_0^2$, $\chi_2^2=\zeta_3^2+\zeta_4^2+\zeta_5^2=\eta_2^2$ and $\zeta_6=0$. Since $|\eta_0|\gg|\eta_1|$, one can approximately ignore the $\eta_1^2\sin^2 k_5$ term.

It is also easy to see that the Weyl surfaces between bands 3 and 4 are given by $\chi_2=0$, i.e., $\zeta_3=\zeta_4=\zeta_5=0$, which is exactly the $k_1$-$k_2$ plane. However, another set of Weyl surfaces are given by $\chi_0=0$, i.e., $\zeta_1=\zeta_2=\sin k_5=0$, which gives the $k_3$-$k_4$ plane in touch with the above Weyl surface (the $k_1$-$k_2$ plane). Such a configuration is unstable against perturbations in 5d.

This touching of Weyl surfaces is removed when one adds the $\eta_3$ term. Via a perturbation analysis, one can show the $\eta_3$ term splits the above two kinds of Weyl surfaces in the $k_5$ direction for about a distance of order $|\eta_3E/\eta_0\eta_1|$ away.

The Weyl surfaces between bands 4 and 5 and between bands 5 and 6 can then be plotted according to the expression of the functions $\zeta_i$, which are as illustrated in Fig. \ref{TINI}. In particular, the condition $(t_1/t_2)^2+\eta_0^2<1$ limits the number of Weyl surfaces between bands 4 and 5 to only four.


\begin{thebibliography}{26}
\expandafter\ifx\csname natexlab\endcsname\relax\def\natexlab#1{#1}\fi
\expandafter\ifx\csname bibnamefont\endcsname\relax
  \def\bibnamefont#1{#1}\fi
\expandafter\ifx\csname bibfnamefont\endcsname\relax
  \def\bibfnamefont#1{#1}\fi
\expandafter\ifx\csname citenamefont\endcsname\relax
  \def\citenamefont#1{#1}\fi
\expandafter\ifx\csname url\endcsname\relax
  \def\url#1{\texttt{#1}}\fi
\expandafter\ifx\csname urlprefix\endcsname\relax\def\urlprefix{URL }\fi
\providecommand{\bibinfo}[2]{#2}
\providecommand{\eprint}[2][]{\url{#2}}

\bibitem[{\citenamefont{Qi and Zhang}(2011)}]{Qi2011}
\bibinfo{author}{\bibfnamefont{X.-L.} \bibnamefont{Qi}} \bibnamefont{and}
  \bibinfo{author}{\bibfnamefont{S.-C.} \bibnamefont{Zhang}},
  \bibinfo{journal}{Rev. Mod. Phys.} \textbf{\bibinfo{volume}{83}},
  \bibinfo{pages}{1057} (\bibinfo{year}{2011}).

\bibitem[{\citenamefont{Berry}(1984)}]{Berry1984}
\bibinfo{author}{\bibfnamefont{M.~V.} \bibnamefont{Berry}},
  \bibinfo{journal}{Proc. R. Soc. Lond. A} \textbf{\bibinfo{volume}{392}},
  \bibinfo{pages}{45} (\bibinfo{year}{1984}), ISSN \bibinfo{issn}{0080-4630}.

\bibitem[{\citenamefont{Nielsen and Ninomiya}(1983)}]{Nielsen1983}
\bibinfo{author}{\bibfnamefont{H.}~\bibnamefont{Nielsen}} \bibnamefont{and}
  \bibinfo{author}{\bibfnamefont{M.}~\bibnamefont{Ninomiya}},
  \bibinfo{journal}{Phys. Lett. B} \textbf{\bibinfo{volume}{130}},
  \bibinfo{pages}{389} (\bibinfo{year}{1983}).

\bibitem[{\citenamefont{Wan et~al.}(2011)\citenamefont{Wan, Turner, Vishwanath,
  and Savrasov}}]{Wan2011}
\bibinfo{author}{\bibfnamefont{X.}~\bibnamefont{Wan}},
  \bibinfo{author}{\bibfnamefont{A.~M.} \bibnamefont{Turner}},
  \bibinfo{author}{\bibfnamefont{A.}~\bibnamefont{Vishwanath}},
  \bibnamefont{and} \bibinfo{author}{\bibfnamefont{S.~Y.}
  \bibnamefont{Savrasov}}, \bibinfo{journal}{Phys. Rev. B}
  \textbf{\bibinfo{volume}{83}}, \bibinfo{pages}{205101}
  (\bibinfo{year}{2011}).

\bibitem[{\citenamefont{Balents}(2011)}]{Balents2011}
\bibinfo{author}{\bibfnamefont{L.}~\bibnamefont{Balents}},
  \bibinfo{journal}{Physics} \textbf{\bibinfo{volume}{4}}, \bibinfo{pages}{36}
  (\bibinfo{year}{2011}).

\bibitem[{\citenamefont{Fu et~al.}(2007)\citenamefont{Fu, Kane, and
  Mele}}]{Fu2007}
\bibinfo{author}{\bibfnamefont{L.}~\bibnamefont{Fu}},
  \bibinfo{author}{\bibfnamefont{C.~L.} \bibnamefont{Kane}}, \bibnamefont{and}
  \bibinfo{author}{\bibfnamefont{E.~J.} \bibnamefont{Mele}},
  \bibinfo{journal}{Phys. Rev. Lett.} \textbf{\bibinfo{volume}{98}},
  \bibinfo{pages}{106803} (\bibinfo{year}{2007}).

\bibitem[{\citenamefont{Qi et~al.}(2008)\citenamefont{Qi, Hughes, and
  Zhang}}]{Qi2008}
\bibinfo{author}{\bibfnamefont{X.-L.} \bibnamefont{Qi}},
  \bibinfo{author}{\bibfnamefont{T.~L.} \bibnamefont{Hughes}},
  \bibnamefont{and} \bibinfo{author}{\bibfnamefont{S.-C.} \bibnamefont{Zhang}},
  \bibinfo{journal}{Phys. Rev. B} \textbf{\bibinfo{volume}{78}},
  \bibinfo{pages}{195424} (\bibinfo{year}{2008}).

\bibitem[{\citenamefont{Murakami}(2007)}]{Murakami2007}
\bibinfo{author}{\bibfnamefont{S.}~\bibnamefont{Murakami}},
  \bibinfo{journal}{New J. Phys.} \textbf{\bibinfo{volume}{9}},
  \bibinfo{pages}{356} (\bibinfo{year}{2007}).

\bibitem[{\citenamefont{Murakami et~al.}(2016)\citenamefont{Murakami, Hirayama,
  Okugawa, and Miyake}}]{Murakami2016}
\bibinfo{author}{\bibfnamefont{S.}~\bibnamefont{Murakami}},
  \bibinfo{author}{\bibfnamefont{M.}~\bibnamefont{Hirayama}},
  \bibinfo{author}{\bibfnamefont{R.}~\bibnamefont{Okugawa}}, \bibnamefont{and}
  \bibinfo{author}{\bibfnamefont{T.}~\bibnamefont{Miyake}},
  \bibinfo{journal}{ArXiv e-prints}  (\bibinfo{year}{2016}),
  \eprint{1610.07132}.

\bibitem[{\citenamefont{Kohmoto et~al.}(1992)\citenamefont{Kohmoto, Halperin,
  and Wu}}]{Kohmoto1992}
\bibinfo{author}{\bibfnamefont{M.}~\bibnamefont{Kohmoto}},
  \bibinfo{author}{\bibfnamefont{B.~I.} \bibnamefont{Halperin}},
  \bibnamefont{and} \bibinfo{author}{\bibfnamefont{Y.-S.} \bibnamefont{Wu}},
  \bibinfo{journal}{Phys. Rev. B} \textbf{\bibinfo{volume}{45}},
  \bibinfo{pages}{13488} (\bibinfo{year}{1992}).

\bibitem[{\citenamefont{Haldane}(2004)}]{Haldane2004}
\bibinfo{author}{\bibfnamefont{F.~D.~M.} \bibnamefont{Haldane}},
  \bibinfo{journal}{Phys. Rev. Lett.} \textbf{\bibinfo{volume}{93}},
  \bibinfo{pages}{206602} (\bibinfo{year}{2004}).

\bibitem[{\citenamefont{Ho\ifmmode~\check{r}\else
  \v{r}\fi{}ava}(2005)}]{Horava2005}
\bibinfo{author}{\bibfnamefont{P.}~\bibnamefont{Ho\ifmmode~\check{r}\else
  \v{r}\fi{}ava}}, \bibinfo{journal}{Phys. Rev. Lett.}
  \textbf{\bibinfo{volume}{95}}, \bibinfo{pages}{016405}
  (\bibinfo{year}{2005}),
  \urlprefix\url{http://link.aps.org/doi/10.1103/PhysRevLett.95.016405}.

\bibitem[{\citenamefont{Zhao and Wang}(2013)}]{Zhao2013}
\bibinfo{author}{\bibfnamefont{Y.~X.} \bibnamefont{Zhao}} \bibnamefont{and}
  \bibinfo{author}{\bibfnamefont{Z.~D.} \bibnamefont{Wang}},
  \bibinfo{journal}{Phys. Rev. Lett.} \textbf{\bibinfo{volume}{110}},
  \bibinfo{pages}{240404} (\bibinfo{year}{2013}),
  \urlprefix\url{http://link.aps.org/doi/10.1103/PhysRevLett.110.240404}.

\bibitem[{\citenamefont{Schnyder and Brydon}(2015)}]{Schnyder2015}
\bibinfo{author}{\bibfnamefont{A.~P.} \bibnamefont{Schnyder}} \bibnamefont{and}
  \bibinfo{author}{\bibfnamefont{P.~M.~R.} \bibnamefont{Brydon}},
  \bibinfo{journal}{J. Phys.: Condens. Matter} \textbf{\bibinfo{volume}{27}},
  \bibinfo{pages}{243201} (\bibinfo{year}{2015}),
  \urlprefix\url{http://stacks.iop.org/0953-8984/27/i=24/a=243201}.

\bibitem[{\citenamefont{Lian and Zhang}(2016)}]{Lian2016}
\bibinfo{author}{\bibfnamefont{B.}~\bibnamefont{Lian}} \bibnamefont{and}
  \bibinfo{author}{\bibfnamefont{S.-C.} \bibnamefont{Zhang}},
  \bibinfo{journal}{Phys. Rev. B} \textbf{\bibinfo{volume}{94}},
  \bibinfo{pages}{041105} (\bibinfo{year}{2016}).

\bibitem[{\citenamefont{Mathai and Thiang}(2017)}]{Mathai2017}
\bibinfo{author}{\bibfnamefont{V.}~\bibnamefont{Mathai}} \bibnamefont{and}
  \bibinfo{author}{\bibfnamefont{G.~C.} \bibnamefont{Thiang}},
  \bibinfo{journal}{J. Phys. A: Math. Theor.} \textbf{\bibinfo{volume}{50}},
  \bibinfo{pages}{11LT01} (\bibinfo{year}{2017}),
  \urlprefix\url{http://stacks.iop.org/1751-8121/50/i=11/a=11LT01}.

\bibitem[{\citenamefont{{Mathai} and {Thiang}}(2016)}]{Mathai2016b}
\bibinfo{author}{\bibfnamefont{V.}~\bibnamefont{{Mathai}}} \bibnamefont{and}
  \bibinfo{author}{\bibfnamefont{G.~C.} \bibnamefont{{Thiang}}},
  \bibinfo{journal}{ArXiv e-prints}  (\bibinfo{year}{2016}),
  \eprint{1611.08961}.

\bibitem[{\citenamefont{Sugawa et~al.}(2016)\citenamefont{Sugawa,
  Salces-Carcoba, Perry, Yue, and Spielman}}]{Sugawa2016}
\bibinfo{author}{\bibfnamefont{S.}~\bibnamefont{Sugawa}},
  \bibinfo{author}{\bibfnamefont{F.}~\bibnamefont{Salces-Carcoba}},
  \bibinfo{author}{\bibfnamefont{A.~R.} \bibnamefont{Perry}},
  \bibinfo{author}{\bibfnamefont{Y.}~\bibnamefont{Yue}}, \bibnamefont{and}
  \bibinfo{author}{\bibfnamefont{I.~B.} \bibnamefont{Spielman}},
  \bibinfo{journal}{ArXiv e-prints}  (\bibinfo{year}{2016}),
  \eprint{1610.06228}.

\bibitem[{\citenamefont{Yang}(1978)}]{Yang1978}
\bibinfo{author}{\bibfnamefont{C.~N.} \bibnamefont{Yang}}, \bibinfo{journal}{J.
  Math. Phys.} \textbf{\bibinfo{volume}{19}} (\bibinfo{year}{1978}).

\bibitem[{\citenamefont{Wilczek and Zee}(1984)}]{Wilczek1984}
\bibinfo{author}{\bibfnamefont{F.}~\bibnamefont{Wilczek}} \bibnamefont{and}
  \bibinfo{author}{\bibfnamefont{A.}~\bibnamefont{Zee}},
  \bibinfo{journal}{Phys. Rev. Lett.} \textbf{\bibinfo{volume}{52}},
  \bibinfo{pages}{2111} (\bibinfo{year}{1984}).

\bibitem[{\citenamefont{Zhang and Hu}(2001)}]{Zhang2001}
\bibinfo{author}{\bibfnamefont{S.-C.} \bibnamefont{Zhang}} \bibnamefont{and}
  \bibinfo{author}{\bibfnamefont{J.}~\bibnamefont{Hu}},
  \textbf{\bibinfo{volume}{294}}, \bibinfo{pages}{823} (\bibinfo{year}{2001}),
  ISSN \bibinfo{issn}{0036-8075},
  \urlprefix\url{http://science.sciencemag.org/content/294/5543/823}.

\bibitem[{\citenamefont{Kitaev}(2009)}]{Kitaev2009}
\bibinfo{author}{\bibfnamefont{A.}~\bibnamefont{Kitaev}}, \bibinfo{journal}{AIP
  Conference Proceedings} \textbf{\bibinfo{volume}{1134}}
  (\bibinfo{year}{2009}).

\bibitem[{\citenamefont{Ryu et~al.}(2010)\citenamefont{Ryu, Schnyder, Furusaki,
  and Ludwig}}]{Ryu2010}
\bibinfo{author}{\bibfnamefont{S.}~\bibnamefont{Ryu}},
  \bibinfo{author}{\bibfnamefont{A.~P.} \bibnamefont{Schnyder}},
  \bibinfo{author}{\bibfnamefont{A.}~\bibnamefont{Furusaki}}, \bibnamefont{and}
  \bibinfo{author}{\bibfnamefont{A.~W.~W.} \bibnamefont{Ludwig}},
  \bibinfo{journal}{New J. Phys.} \textbf{\bibinfo{volume}{12}},
  \bibinfo{pages}{065010} (\bibinfo{year}{2010}),
  \urlprefix\url{http://stacks.iop.org/1367-2630/12/i=6/a=065010}.

\bibitem[{\citenamefont{Wigner}(1932)}]{Wigner1932}
\bibinfo{author}{\bibfnamefont{E.~P.} \bibnamefont{Wigner}},
  \bibinfo{journal}{Nachr. Akad. Wiss. G\"{o}ttingen, Math. Physik. Kl.}
  \textbf{\bibinfo{volume}{546}} (\bibinfo{year}{1932}).

\bibitem[{\citenamefont{Dyson}(1962)}]{Dyson1962}
\bibinfo{author}{\bibfnamefont{F.~J.} \bibnamefont{Dyson}},
  \bibinfo{journal}{J. Math. Phys.} \textbf{\bibinfo{volume}{3}},
  \bibinfo{pages}{1199} (\bibinfo{year}{1962}).

\bibitem[{\citenamefont{Georgi and Glashow}(1972)}]{Georgi1972}
\bibinfo{author}{\bibfnamefont{H.}~\bibnamefont{Georgi}} \bibnamefont{and}
  \bibinfo{author}{\bibfnamefont{S.~L.} \bibnamefont{Glashow}},
  \bibinfo{journal}{Phys. Rev. Lett.} \textbf{\bibinfo{volume}{28}},
  \bibinfo{pages}{1494} (\bibinfo{year}{1972}),
  \urlprefix\url{http://link.aps.org/doi/10.1103/PhysRevLett.28.1494}.

\end{thebibliography}

\end{document}